\documentclass[prl,nofootinbib,twocolumn
]{revtex4-2}

\usepackage{amsmath,amssymb,bm,graphicx}
\usepackage{multirow}% http://ctan.org/pkg/multirow

\usepackage{dcolumn}% Align table columns on decimal point
\usepackage{latexsym}
\usepackage[dvipdfmx, colorlinks=true, linkcolor=blue]{hyperref} 
\def\<{\langle}
\def\>{\rangle}
\def\oper{{\mathchoice{\rm 1\mskip-4mu l}{\rm 1\mskip-4mu l}
{\rm 1\mskip-4.5mu l}{\rm 1\mskip-5mu l}}}

\newcommand{\tr}{\mathrm{Tr}}

\newcommand{\LA}{\mathop{\mathcal{L}}\nolimits}

\newcommand{\bracket}[1]{\langle #1  \rangle}
\newcommand{\BH}{\mathcal{B}(\mathcal{H})}

\newtheorem{Theorem}{Theorem}
\newtheorem{Proposition}{Proposition}

\newtheorem{Lemma}{Lemma}

\newtheorem{CON}{Conjecture}

\begin{document}
%\title{\textbf{Non-Markovian quantum evolution from the dynamics of classical-quantum states}}
\title{\textbf{On the universal constraints for relaxation rates for quantum dynamical semigroup}}%: classical vs. quantum}}

%\title{\textbf{Tight constraints on relaxation times for quantum dynamical semigroup}}
\author{Dariusz Chru\'sci\'nski$^1$, Gen Kimura$^2$, Andrzej Kossakowski$^1$, Yasuhito Shishido$^2$}
\affiliation{$^1$Institute of Physics, Faculty of Physics, Astronomy and Informatics  Nicolaus Copernicus University, Grudzi\c{a}dzka 5/7, 87--100 Toru\'n, Poland \\
$^2$ College of Systems Engineering and Science,
Shibaura Institute of Technology, Saitama 330-8570, Japan
}

%\affiliation{Institute of Physics, Faculty of Physics, Astronomy and Informatics, \\ Nicolaus Copernicus University,
%Grudzi\c{a}dzka 5/7, 87--100 Toru\'n, Poland\\
%}


\begin{abstract}
%\re{In order to analyze a generic behavior of quantum Markovian dynamics,
A conjecture for the universal constraints for relaxation rates of a quantum dynamical semigroup is proposed. It is shown that it holds for several interesting classes of semigroups, e.g.  unital semigroups and semigroups derived in the weak coupling limit from the proper microscopic model. Moreover, proposed conjecture  is supported by numerical analysis. This conjecture has several important implications: it allows to provide universal constraints for spectra of quantum channels and provides necessary condition to decide whether a given channel is consistent with Markovian evolution.
\end{abstract}

\maketitle

%======SECTION=================INTRODUCTION=================
%\section{Introduction}

{\em Introduction} --- Spectral analysis belongs to the heart of quantum theory \cite{von-Neumann}. Actually, this is {\em spectroscopy} which gave birth to quantum theory. Very often one infers information about the quantum system  measuring a spectrum of some operator representing physical objects (quantum observables, quantum maps, etc.). In this Letter we analyze the spectral properties of the celebrated Gorini-Kossakowski-Lindblad-Sudarshan (GKLS) generator of quantum Markovian semigroup \cite{GKS,L}
\begin{equation}\label{ME}
  \dot{\rho} = \mathcal{L}(\rho) ,
\end{equation}
where $\mathcal{L}$ has the following well known form
\begin{equation}\label{L}
\mathcal{L}(\rho)= - i [H,\rho] + \sum_{k} \gamma_k \left(L_k\rho L_k^\dagger -\frac 12\{L_k^\dagger L_k,\rho\}\right) ,
\end{equation}
with arbitrary noise operators $L_k$ and positive rates $\gamma_k$.
This is the most general structure of the generator which guaranties that the dynamical map $\Lambda_t = e^{t \mathcal{L}}$ is completely positive and trace-preserving (CPTP) \cite{GKS,L,Alicki,Breuer}. Solutions of  (\ref{ME}) define very good approximation of real system's evolution provided the system-environment interaction is sufficiently weak and there is separation of time scales for the system  and environment \cite{Breuer}. Typical examples where Markovian approximation is physically justified are quantum optical systems \cite{Gardiner,Plenio,Car}. It is well known that eigenvalues of $\mathcal{L}$ provide information about the rate of relaxation, dissipation and decoherence processes and hence define key physical property of the physical process. Actually, these are not $\gamma_k$ which are directly measured in the laboratory but the corresponding eigenvalues of the generator.

%A generator defines a linear map $\mathcal{L} : \BH \to \BH$ and hence one can apply standard tools of linear algebra asking for eigenvalues and eigenvectors.

Let $\ell_\alpha$ be the corresponding (complex) eigenvalues of $\mathcal{L}$, that is, $\mathcal{L}(X_\alpha) = \ell_\alpha X_\alpha$ for $\alpha=0,\ldots,d^2-1$, where $d = {\rm dim}\, \mathcal{H}$. Since $\mathcal{L}$ does preserve Hermiticity one has $\mathcal{L}(X^\dagger_\alpha) = \ell^*_\alpha X_\alpha^\dagger$, that is, if $\ell_\alpha$ is complex, then $\ell^*_\alpha$ is also an eigenvalue. It is well known \cite{Alicki} that $\lambda_0=0$ and the corresponding eigenvector (zero-mode of $\mathcal{L}$) $X_0$ gives rise to the invariant state of the evolution $\omega=X_0/{\rm Tr}\,X_0$, that is, $\Lambda_t(\omega)=\omega$. The corresponding eigenvalues $\lambda_\alpha(t)$ of the dynamical map $\Lambda_t = e^{t \mathcal{L}}$ read $\lambda_\alpha(t) = e^{t \ell_\alpha}$ and hence necessarily the relaxation rates $ \Gamma_\alpha$ defined by

\begin{equation}\label{}
  \Gamma_\alpha = - {\rm Re}\, \ell_\alpha ,
\end{equation}
are non-negative $\Gamma_\alpha \geq 0$ for all $\alpha=1,\ldots,d^2-1$ (otherwise $e^{t \ell_\alpha}$ blows up as $t \to \infty$). Eigenvalues $\lambda_\alpha(t)$ of the corresponding dynamical map $\Lambda_t = e^{t\mathcal{L}}$ belong to the unit disc on the complex plane, that is, $|\lambda_\alpha(t)|\leq 1$. This is a quantum analog of the celebrated Frobenius-Perron theorem for stochastic matrices. Surprisingly, apart form the fact that all $\Gamma_\alpha \geq 0$ not much more is known about the structure of the spectrum of a GKSL generator. Actually, one can show that $e^{t \mathcal{L}}$ is CPTP for $t\geq 0$ if and only if $\mathcal{L}$ satisfy the following property (known as conditional complete positivity) \cite{Wolf-Isert,WOLF}

\begin{equation}\label{PLP}
  P^\perp [(I \otimes \mathcal{L})(P)] P^\perp \geq 0 ,
\end{equation}
where $P=|\psi_+\>\<\psi_+|$ denotes {the} projector  onto maximally mixed state $|\psi_+\> \in \mathcal{H} \otimes \mathcal{H}$, and $P^\perp = \oper - P$ is orthogonal to $P$. Unfortunately, condition (\ref{PLP}) does not provide any transparent information about the spectrum of $\mathcal{L}$. The same problem arises for quantum channels. A linear map $\Phi : \BH \to \BH$ is completely positive if and {only} if the corresponding Choi matrix $(I \otimes \Phi)(P) \geq 0$ \cite{Choi}. Again, positivity of the Choi matrix cannot be easily translated into the property of the spectrum of the map $\Phi$. This should be clear since the map and hence also its Choi matrix depend in a  nontrivial way both on the spectrum (eigenvalues) and eigenvectors. On the other hand eigenvalues and in particular relaxation rates have a clear physical interpretation and can be directly measured. Hence, eigenvalues of the Choi matrix decide about complete positivity and eigenvalues of the generator (or the  quantum channel) are measurable quantities. It is, therefore, clear that one can expect some additional property relating relaxation rates which is responsible for complete positivity of the quantum evolution. Relaxation properties of GKLS generators were further studied in \cite{Alicki,Spohn} and more recently e.g. in \cite{Dietz,Heide}. Some constraints for relaxation rates for 3- and 4-level systems were presented in \cite{Solomon,Berman,Schirmer}. Interestingly, authors of a seminal paper \cite{GKS} already observed that for a qubit evolution governed by the following well known generator

\begin{equation}\label{QUBIT}
  \mathcal{L}(\rho) = - i\frac{\Delta}{2}[\sigma_z,\rho] + \mathcal{L}_D(\rho)  ,
\end{equation}
with the dissipative part $\mathcal{L}_D = \gamma_+ \mathcal{L}_+ + \gamma_- \mathcal{L}_- + \gamma_z \mathcal{L}_z$ consisting of: pumping $\mathcal{L}_+(\rho) = \sigma_+ \rho \sigma_- - \frac 12\{\sigma_-\sigma_+,\rho\}$, damping $\mathcal{L}_-(\rho) = \sigma_- \rho \sigma_+ - \frac 12\{\sigma_+\sigma_-,\rho\}$, and dephasing $\mathcal{L}_z(\rho) = \sigma_z \rho \sigma_z - \rho$, complete positivity implies the following well known condition for the relaxation times $T_\alpha = 1/\Gamma_\alpha$:

\begin{equation}\label{TT}
  T_{\rm L} \geq 2\, T_{\rm T} ,
\end{equation}
where the longitudinal rate $\Gamma_{\rm L} = \Gamma_3= \gamma_+ + \gamma_-$, and transversal rate $\Gamma_{\rm T} = \Gamma_1=\Gamma_2 = \frac{1}{2} (\gamma_+ + \gamma_- ) + \gamma_z$. Condition (\ref{TT}) was experimentally demonstrated to be true \cite{Alicki,LT}. Clearly, the very condition (\ref{TT}) provides only partial information about the corresponding qubit generator. However, violation of (\ref{TT}) shows that the generator does not provide legitimate CPTP evolution. Condition (\ref{TT}) has even more appealing form when rephrased in terms of relaxation rates. Indeed, one finds

\begin{equation}\label{GGG}
  \sum_{k=1}^3 \Gamma_k \geq 2 \Gamma_i \ ; \ \ \ i=1,2,3 ,
\end{equation}
that is, each single relaxation rate cannot be too large. In terms of relative relaxation rates $R_i = \Gamma_i/(\Gamma_1+\Gamma_2+\Gamma_3)$, it says that,

\begin{equation}\label{Ri}
  R_i \leq \frac 12 \ ; \ \ \ i=1,2,3 ,
\end{equation}
The generator (\ref{QUBIT}) is very special and in particular implies that the rates $\Gamma_1$ and $\Gamma_2$ are the same. Interestingly, Kimura \cite{GK} showed that condition (\ref{GGG}) is universal for any qubit generator. For a purely dissipative generator Wolf and Cirac derived the following result (Theorem 6 in \cite{Wolf-Cirac})

\begin{equation}\label{}
  \| \mathcal{L} \| \leq \frac 2d \Gamma ,
\end{equation}
with $\Gamma := \sum_{\beta=1}^{d^2-1} \Gamma_\beta$, {where $ \| \mathcal{L} \|$ denotes the operator norm}.
Note, that due to $ \| \mathcal{L} \| \geq |\ell_\alpha| \geq \Gamma_\alpha$, the above condition implies
\begin{equation}\label{R-Wolf}
  R_\alpha \leq \frac 2d \ ; \ \ \ \alpha=1,\ldots,d^2-1 ,
\end{equation}
where $R_\alpha =  \Gamma_\alpha/\Gamma$. Recently, Kimura et al. \cite{KAW} {obtained the following universally valid constraints for any GKLS generator:}
\begin{equation}\label{Gen}
  R_\alpha \leq \frac{\sqrt{2}}{d} \ ; \ \ \ \alpha=1,\ldots,d^2-1.
\end{equation}
In this Letter we conjecture that the bound (\ref{Gen}) can be still improved and propose the following

\begin{CON} \label{CON-1} Any GKLS generator \eqref{L} for $d$-level quantum systems implies the following constraints for the relaxation rates

\begin{equation}\label{CON}
  \Gamma  \geq d \Gamma_\alpha \ ; \ \ \ \alpha=1,\ldots,d^2-1 .
\end{equation}
Equivalently, in terms of the relative relaxation rates $R_\alpha = \Gamma_\alpha/\Gamma$, we conjecture that

\begin{equation}\label{Rid}
  R_\alpha \leq \frac 1d \ ; \ \ \ \alpha=1,\ldots,d^2-1.
\end{equation}
{Moreover, the bound (\ref{CON}) is tight, i.e. cannot be improved.}
\end{CON}
Unfortunately, we still do not have a complete proof of (\ref{CON}). However, we show in this Letter that this conjecture holds for several important classes of GKLS generators. In particular any generator giving rise to the unital evolution, that is $\mathcal{L}(\oper)=0$, satisfies (\ref{CON}). Unital (often called doubly stochastic) maps {characterize decoherence processes that does not decrease entropy \cite{entropy1,entropy2}} and provide direct generalization of unitary maps.
A second important class are GKLS generator which display additional symmetry, that is, they are covariant w.r.t. maximal abelian subgroup of the unitary group $U(d)$. Actually, qubit generator (\ref{QUBIT}) belongs to this class. The classical Pauli master equation is another example.

\begin{figure}[t]
\includegraphics[width=9cm]{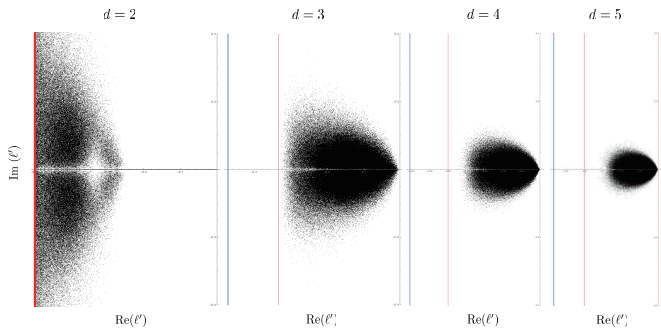}
\caption{Distributions of eigenvalues of random Lindbladians.
For each $d=2,3,4,5$, we randomly generated $100,000$ GKLS generators and plotted the normalized eigenvalues $\ell' := \ell_\alpha/\Gamma$. Red vertical lines denote the bound `$-1/d$', corresponding to our conjecture, while blue ones denote the previously obtained bound $-\sqrt{2}/d$ \cite{KAW}. }\label{fig:ns1}
\end{figure}

The formula  (\ref{L}) provides the most general mathematical structure of the generator compatible with the requirement of complete positivity and trace-preservation. Note, however, that not every generator constructed according to (\ref{L}) has a clear physical interpretation. There exists a natural class of generators of Markovian semigroups derived in the weak coupling limit \cite{Davies,Alicki,Breuer} and these do enjoy the covariance property. Hence, we may summarise that physically {motivated} generators do satisfy Conjecture (\ref{CON-1}). This conjecture is also strongly supported by numerical analysis (cf. Figure \ref{fig:ns1}).

Interestingly, it is perfectly consistent with the spectrum of  random GKLS generator  in the large $d$-limit \cite{Denisov}. {Finally, we also construct a GKLS model which saturates \eqref{Rid} (for some $\alpha$). This implies that \eqref{Rid} are the tightest constraints which characterize the universally valid spectral property of GKLS generators. }

Clearly, the conjecture providing universal constraints for relaxation rates is interesting by itself {since they are composed of experimentally accessible quantities and hence provide a direct method to check the validity of GKLS generators, or the completely positive condition. }
It has, however,  further very interesting implications. It allows to establish universal constraints  for eigenvalues of quantum channels (Conjecture \ref{CON-2}).  Moreover, it provides necessary condition for a quantum channel $\Phi$ to be represented via $\Phi = e^\mathcal{L}$ for some GKLS generator \cite{Wolf-Isert}. It is found that in this case all eigenvalues are constrained to a ring $r \leq |z_\alpha| \leq 1$, where the inner radius $r$ is fully characterized by the original channel $\Phi$ (Conjecture \ref{CON-3}).

\vspace{.2cm}

\noindent {\em Classical Pauli master equation.} --- Let us start our analysis with a classical counterpart of master equation. Consider a Pauli rate equation for a classical system with $d$ states

\begin{equation}\label{PME}
  \frac{dp_i}{dt} = \sum_{j =1}^d  K_{ij} p_j ,
\end{equation}
where $K$ is the classical generator satisfying the following Kolmogorov conditions \cite{Kampen}
\begin{equation}\label{KOL}
  K_{ij} \geq 0 \ , \ (i\neq j) \ \ ; \ \  \sum_{i=1}^d K_{ij}= 0 .
\end{equation}
Hence $K_{ij}$ can be represented as $ K_{ij} = t_{ij} - \delta_{ij} \sum_{m=1}^d t_{mj}$, with $t_{ij} \geq 0$.
{Note that here only $t_{ij}$ with $i \neq j$ are relevant, so in the following, we put $t_{ii} =0$. }
Equivalently, (\ref{PME}) can be formulated as follows

\begin{equation}\label{PMEt}
  \dot{p}_i= \sum_{j =1}^d (t_{ij} p_j - t_{ji} p_i) .
\end{equation}
Do we have a classical analog of (\ref{CON})?  Spectral properties of $d \times d$ matrix $K_{ij}$ are similar to that of $\mathcal{L}$: there are $d$ complex eigenvalues {$\ell^{\rm cl}_0,\ldots,\ell^{\rm cl}_{d-1}$} with $\ell^{\rm cl}_0=0$. Moreover,  $\Gamma^{\rm cl}_k = - {\rm Re}\,  \ell^{\rm cl}_k \geq 0$, and the spectrum is symmetric w.r.t. real axis.  Interestingly, in the classical case there is no bound on the relative classical rates $R^{\rm cl}_k$, that is, given a set of classical rates $\Gamma^{\rm cl}_k \geq 0$ one can construct a classical generator $K_{ij}$ which does display exactly these rates. In particular any  single relative rate $R^{\rm cl}_k$ can be arbitrary close to `$1$'. (See Appendix \ref{sec:NoBddCl} for details.)

Consider now a quantum evolution $t \to \rho(t)$ such that the diagonal elements $p_k = \rho_{kk}$ evolve according to the classical Pauli equation (\ref{PME}). Introducing the family of noise operators $E_{ij} = |i\>\<j|$ one constructs the following GKLS generator

\begin{equation}\label{L-class}
 \LA(\rho) = \sum_{i,j=1}^d t_{ij} E_{ij} \rho E_{ij}^\dagger  - \frac 12 \{ B  ,\rho\}\,
\end{equation}
where $B= \sum_k b_k |k\>\<k|$, with $b_k = \sum_{j=1}^d t_{jk}$.
The spectrum of $\LA$ consists of $d$ {\em classical} eigenvalues of the classical generator represented by the matrix $K_{ij} = t_{ij} - \delta_{ij} b_j$:  $\lambda_0=0,\ell^{\rm cl}_1,\ldots,\ell^{\rm cl}_{d-1}$, and the remaining eigenvalues correspond to eigenvectors $E_{kl}$:

\begin{equation}\label{}
  \LA(L_{kl}) = - \frac 12(b_k + b_l) E_{kl} \ , \ \ \ (k \neq l) .
\end{equation}
Hence, one has {\em classical} rates $\Gamma^{\rm cl}_1,\ldots,\Gamma^{\rm cl}_{d-1}$,  and the remaining {\em quantum } rates

\begin{equation}\label{Gkl}
  \Gamma_{kl} = \frac 12 (b_k + b_l) , \ \ \ (k \neq l) .
\end{equation}

\begin{Proposition}\label{PrCl} The generator \eqref{L-class} satisfies  \eqref{CON}.
\end{Proposition}
For the proof see Appendix \ref{sec:PrPr1}. This simple analysis shows that the role of {\em quantum} rates $ \Gamma_{kl} $ is to restore the bound (\ref{CON}) which is violated if one considers only {\em classical rates} $\Gamma^{\rm cl}_k$. In terms of relative rates for the original classical problem $R^{\rm cl}_k $ can be arbitrarily close to `$1$'. However, after incorporating the remaining rates $\Gamma_{kl}$ one finds

$$   R^{\rm cl}_k \leq \frac 1d \ , \ \ \ R_{kl} \leq \frac 1d . $$
Clearly, this is the requirement of complete positivity which enforces the rates to satisfy (\ref{CON}).

\vspace{.2cm}

\noindent  {\em The bound is tight.} ---  For any dimension $d$ one can construct $\mathcal{L}$ such that bound (\ref{Rid}) is attained for some $R_\alpha$. Indeed, consider well known generator constructed via a double commutator

\begin{equation}\label{pure}
  \mathcal{L}(\rho)  = -  [\Sigma,[\Sigma,\rho]] = 2\Sigma \rho  \Sigma -  \{\Sigma^2,\rho \}  ,
\end{equation}
for some Hermitian operator $\Sigma$. A well known example is a qubit dephasing corresponding to $\Sigma=\sigma_z$. Let $\Sigma = \sum_k s_k |k\>\<k|$ and assume that $s_1 \leq \ldots \leq s_d$.  Then one finds for the relaxation rates $\Gamma_{ij} = (s_i-s_j)^2$ with the maximal rate $\Gamma_{\rm max} = \Gamma_{1d}$. One shows (cf. Appendix \ref{sec:TightBdd}) that

\begin{equation}\label{}
  \sum_{i,j=1}^d \Gamma_{ij} \geq d \Gamma_{\rm max} ,
\end{equation}
which supports the conjecture (\ref{CON}). Moreover, taking $  s_2 = \ldots = s_{d-1} = \frac{s_1+s_d}{2}$, one finds $\sum_{i,j=1}^d \Gamma_{ij} = d \Gamma_{\rm max}$, or equivalently $R_{\rm max} = \frac 1d$.

\vspace{.2cm}

\noindent {\em Dissipativity condition.} --- It is more convenient to proceed in the Heisenberg picture  defined by the dual generator $\mathcal{L}^\ddag$ which is related to Schr\"odinger picture generator $\mathcal{L}$ via ${\rm Tr}(X \mathcal{L}(Y)) = {\rm Tr}(\mathcal{L}^\ddag(X)Y)$ for any pair of operators $X,Y \in \BH$. Clearly, both $\mathcal{L}$ and $\mathcal{L}^\ddag$ have the same spectrum $\ell_\alpha$ but in general different eigenvectors.
As was shown by Lindblad \cite{L} any GKLS generator satisfy the following dissipativity condition

\begin{equation}\label{1}
  \mathcal{L}^\ddag(X^\dagger X) - \mathcal{L}^\ddag(X^\dagger)X- X^\dagger \mathcal{L}^\ddag(X) \geq 0 ,
\end{equation}
for all $X \in \BH$. Inserting the formula (\ref{L}) for the generator one finds (cf. Appendix \ref{sec:Derivation(23)})

\begin{equation}\label{2}
  \mathcal{L}^\ddag(X^\dagger X) - \mathcal{L}^\ddag(X^\dagger)X- X^\dagger \mathcal{L}^\ddag(X)  = \sum_k \gamma_k [L_k,X]^\dagger [L_k,X] .
\end{equation}
Now, inserting $X = Y_\alpha$, where $\mathcal{L}^\ddag(Y_\alpha) = \ell_\alpha Y_\alpha$ one obtains

\begin{equation*}
  \mathcal{L}^\ddag(Y_\alpha^\dagger Y_\alpha) + {2} \Gamma_\alpha Y_\alpha^\dagger Y_\alpha  = \sum_k \gamma_k [L_k,Y_\alpha]^\dagger [L_k,Y_\alpha] ,
\end{equation*}
which finally implies

\begin{eqnarray*}
 && {\rm Tr}(\omega \mathcal{L}^\ddag(Y_\alpha^\dagger Y_\alpha)) + 2 \Gamma_\alpha {\rm Tr}(\omega Y_\alpha^\dagger Y_\alpha) \\
  && \quad = \sum_k \gamma_k {\rm Tr}(\omega [L_k,Y_\alpha]^\dagger [L_k,Y_\alpha] ) ,
\end{eqnarray*}
where $\omega$ is an invariant state satisfying $\mathcal{L}(\omega)=0$. One has therefore

$$  {\rm Tr}(\omega \mathcal{L}^\ddag(Y_\alpha^\dagger Y_\alpha)) = {\rm Tr}(\mathcal{L}(\omega) Y_\alpha^\dagger Y_\alpha) = 0 , $$
and hence one finds the following formula for $\Gamma_\alpha$

\begin{equation}\label{!}
  2 \Gamma_\alpha {\rm Tr}(\omega Y_\alpha^\dagger Y_\alpha)  = \sum_k \gamma_k {\rm Tr}(\omega [L_k,Y_\alpha]^\dagger [L_k,Y_\alpha] ) .
\end{equation}
Introducing the following inner product $(A,B)_\omega =  {\rm Tr}(\omega  A^\dagger B)$ and the corresponding $\omega$-norm $\|A\|_\omega^2 = (A,A)_\omega$ the formula (\ref{!}) may be rewritten in the following compact form

\begin{equation}\label{!a}
   \Gamma_\alpha  = \frac{1}{2 \| Y_\alpha\|^2_\omega}  \sum_k \gamma_k \| [L_k,Y_\alpha] \|^2_\omega .
\end{equation}
This formula is universal, that is, it holds for any GKLS generator. Clearly, to compute $\Gamma_\alpha$ one has to know the corresponding eigenvector $Y_\alpha$ and the invariant state $\omega$. In particular, since $Y_0= \oper$, one recovers $\Gamma_0=0$.

\vspace{.2cm}

\noindent   {\em Unital semigroups.} --- In this section starting from the universal formula (\ref{!a}) we prove (\ref{CON}) for generators of unital semigroup, i.e. semigroups satisfying $e^{t\mathcal{L}}(\oper) = \oper$.
Unital semigroups enjoy several important properties. One  proves \cite{entropy1,entropy2} that $e^{t\mathcal{L}}$ is unital if and only if for any initial state $\rho$ one has
\begin{equation}\label{}
  \frac{d}{dt} S(e^{t \mathcal{L}}(\rho)) \geq 0 ,
\end{equation}
where $S(\rho)$ stands for the von-Neumann entropy (actually it holds also for R\'enyi and Tsallis entropy as well).
The corresponding generator satisfy $\LA(\oper)=0$.
This condition is equivalent to
\begin{equation}\label{LL-LL}
  \sum_k \gamma_k L_k^\dagger L_k =  \sum_k \gamma_k L_k L_k^\dagger .
\end{equation}
In particular it happens when all Lindblad operators $L_k$  are normal ($L_kL_k^\dagger = L_k^\dagger L_k$).

Inserting  $\omega = \oper/d$ into formula (\ref{!a}) one obtains
\begin{equation}\label{!!}
   \Gamma_\alpha  = \frac{1}{2 \| Y_\alpha\|^2}  \sum_k \gamma_k \| [L_k,Y_\alpha] \|^2 ,
\end{equation}
where now $\|A\|^2 = {\rm Tr}(A^\dagger A)$.  To prove (\ref{CON}) we use the following intricate inequality \cite{BW}

\begin{equation}\label{BW}
  \| [A,B]\|^2 \leq 2 \| A \|^2 \|B \|^2 .
\end{equation}
Actually, this inequality was conjectured by B\"ottcher and Wenzel \cite{BW2} in 2005 (see \cite{BW} for more details). A simpler proof can be found in \cite{ref:Aud}. It should be stressed that the bound \eqref{Gen} was shown by the direct use of this inequality as well.

%who already  proved it for $2 \times 2$ real matrices. Then in 2006 it was proved for real $3 \times 3$ matrices and in 2007 for real $d \times d$ matrices (see \cite{BW} for more details).
%Final proof for complex $d \times d$ can be found in \cite{BW}. \re{(See \cite{ref:Aud} for a simpler and conceptually sound proof.)
%We notice .} \bl{This part ``Actually, ... '' could be written more compactly.}
%It is, therefore, clear that (\ref{BW}) is quite intricate mathematical result.
Now, (\ref{BW}) immediately implies
\begin{equation}\label{!!!}
  \Gamma_\alpha   \leq  \sum_k \gamma_k \| L_k\|^2  .
\end{equation}
Assuming the following normalization $\| L_k \|^2  =1$ {as well as the condition $\tr L_k = 0$ without loss of generality}, one shows (cf. Appendix \ref{sec:Derivation(31)}) that
\begin{equation}\label{eq:ID}
\sum_k \gamma_k = \frac 1 d \sum_\alpha \Gamma_\alpha ,
\end{equation}
and hence (\ref{!!!}) reproduces (\ref{CON}).
{Thus, we have shown
\begin{Theorem}\label{prop:US}
The generator with unital semigroup satisfies \eqref{CON}.
\end{Theorem}
}

\vspace{.2cm}

\noindent {\em A class of covariant generators.} --- Symmetry plays a key role in modern physics. In many cases it enables one to simplify the  problem and  often  leads to much deeper understanding and the more elegant mathematical formulation. Let us consider a class of generators covariant w.r.t. the maximal commutative subgroup of the unitary group $U(d)$

\begin{equation}\label{ULU}
  U_\mathbf{x} \LA(X) U^\dagger_\mathbf{x} = \LA(U_\mathbf{x} X U^\dagger_\mathbf{x}) ,
\end{equation}
where $U_\mathbf{x} =  \sum_{k=1}^d e^{-i x_k} |k\>\<k|$, and $\mathbf{x} = (x_1,\ldots,x_d) \in \mathbb{R}^d$. Any generator satisfying (\ref{ULU}) has the following form

\begin{equation}\label{COV}
  \LA = \LA_0 + \LA_1 + \LA_2 ,
\end{equation}
where $ \LA_0(\rho) =  -i[H,\rho]$, together with
\begin{eqnarray}\label{012}
  \LA_1(\rho) &=& \sum_{{i,j=1}}^d t_{ij} E_{ij} \rho E_{ij}^\dagger  - \frac 12 \{ B ,\rho\}\, \\
  \LA_2(\rho) &=&  \sum_{i, j=1}^d d_{ij}  |i\>\<i|  \rho |j\>\<j| - \frac 12 \{ D ,\rho\} , \nonumber
\end{eqnarray}
where the Hamiltonian $H= \sum_i h_i |i\>\<i| $,  $B=\sum_j b_j |j\>\<j|$, with $b_j = \sum_{i} t_{ij}$, and $D = \sum_{i=1}^d d_{ii} |i\>\<i|$ (See Appendix \ref{sec:CovG-1}).
This is GKLS generator iff $t_{ij} \geq 0$ and the Hermitian matrix $[d_{ij}]_{i,j=1}^d$ is positive definite. Clearly, $\LA_1$ is a classical generator considered before and $\LA_2$ adds pure decoherence with respect to the orthonormal basis $|1\>,\ldots,|d\>$. Interestingly, the very condition (\ref{ULU}) implies that $ \LA_\alpha\, \LA_\beta = \LA_\beta\, \LA_\alpha$ for $\alpha=0,1,2$. Hence $\LA_0,\LA_1,\LA_2$ share the same eigenvectors. It is, therefore, clear that eigenvalues of $\mathcal{L}$ are simply sum of eigenvalues of $\LA_\alpha$. Due to this property the analysis of $\mathcal{L}$ leads to the following

\begin{Proposition}\label{prop:CG} The generator \eqref{COV} satisfies  \eqref{CON}.
\end{Proposition}
For the proof see Appendix \ref{sec:CovG-2}.

\vspace{.2cm}

\noindent   {\em Markovian semigroup in the weak coupling limit.} --- Any legitimate generator of CPTP semigroup has a GKSL form (\ref{L}).  However, not every such generator has a clear physical interpretation. If the (open) quantum system is weakly coupled to the environment it was shown by Davies \cite{Davies,DAVIES} that performing so called weak coupling limit one eventually derives Markovian generator which has exactly GKLS form but now has a clear physical meaning being derived from the proper microscopic model (cf. \cite{Breuer,Alicki,Fabio,Rivas}). Actually, if the invariant state $\omega$  has a non-degenerate spectrum (generic situation), then the corresponding generator derived in the weak coupling limit satisfies (\ref{ULU}), and moreover $[\omega,U_\mathbf{x}]=0$, that is, $\omega = \sum_k \omega_k |k\>\<k|$.
Additional property of such generator is a quantum detailed balance condition \cite{Alicki} which in this case reduces to $t_{ik} \omega_k = t_{ki} \omega_i$, which is, however, not essential for (\ref{CON}). Hence we may conclude that a class of physically legitimate GKLS generators defined via weak coupling limit does satisfy (\ref{CON}).

\vspace{.2cm}

\noindent  {\em Implications.} --- Provided our conjecture is true what is it good for? Note, that it enables to characterize spectra of quantum channels. Indeed, if $\Phi$ is a quantum channel (CPTP map), then $\LA(\rho) = \Phi(\rho) - \rho$ defines a legitimate GKLS generator \cite{GKS}.  An example of such generator is just qubit dephasing `$\sigma_z \rho \sigma_z - \rho$'. Now, let $z_\alpha = x_\alpha + i y_\alpha$ denote eigenvalues of $\Phi$. Clearly, they belong to the unit disc $|z_\alpha| \leq 1$ and $z_0=1$. It is therefore clear that Conjecture \ref{CON-1} implies the following
\begin{CON}  \label{CON-2} The spectrum $z_\alpha = x_\alpha + iy_\alpha$ of any quantum channel satisfy

\begin{equation}\label{xx}
\sum_{\beta=1}^{d^2-1} x_\beta \leq d(d-1) - 1 + d x_\alpha ,
\end{equation}
for $\alpha=1,\ldots,d^2-1$.
\end{CON}
Since the Conjecture \ref{CON-1} holds in the qubit case one has

\begin{Proposition} The spectrum $z_\alpha = x_\alpha + iy_\alpha$ of any qubit channel satisfies
\begin{equation}\label{xxx}
  | x_1 \pm x_2| \leq 1 \pm x_3  .
\end{equation}
\end{Proposition}
Indeed, (\ref{xxx}) follows immediately from (\ref{xx}) for $d=2$. In particular for the Pauli channel $\Phi(\rho) = \sum_{\alpha=0}^{3} p_\alpha \sigma_\alpha \rho \sigma_\alpha$ one has $z_k = x_k$, and (\ref{xxx}) are equivalent to the celebrated Fujiwara-Algoet conditions \cite{Algoet}.
{Moreover}, since the Conjecture 1 holds for generators satisfying $\mathcal{L}(\oper) = 0$, one immediately proves
\begin{Proposition} The spectrum  of any unital quantum channel satisfies \eqref{xx}.
\end{Proposition}
A second immediate implication of Conjecture \ref{CON-1} is the problem of deciding whether a given quantum channel $\Phi$  can be represented as $\Phi = e^{\mathcal{L}}$ for some GKLS generator $\mathcal{L}$ \cite{Wolf-Isert,Wolf-Cirac}. Our original Conjecture \ref{CON-1} implies

\begin{CON} \label{CON-3} If $\Phi = e^{\mathcal{L}}$, then the spectrum $z_\alpha$ of $\Phi$ satisfies

\begin{equation}\label{zzz}
  {\rm det}\, \Phi = z_1 \ldots z_{d^2-1} \leq |z_\alpha|^d ,
\end{equation}
for $\alpha = 1,\ldots,d^2-1$.
\end{CON}
Interestingly, it shows that all $z_\alpha$ are not only constrained to the unit Frobenius disc but belong to the ring

\begin{equation}\label{ring}
  \sqrt[d]{ {\rm det}\, \Phi} \leq |z_\alpha| \leq 1 .
\end{equation}
Clearly, Conjecture \ref{CON-3} is satisfied for all qubit channels and all unital channels. In particular for a qubit Pauli channel all eigenvalues $z_\alpha$ are real and hence (\ref{zzz}) reduces to the following simple condition $z_i z_j \leq z_k$,
where $i,j,k\in \{1,2,3\}$ are all different. This condition was recently derived in \cite{Mario,Karol}.

\vspace{.2cm}

\noindent  {\em Conclusions.} --- In this Letter we propose a conjecture for the universal constraints for relaxation rates $\Gamma_\alpha$ of a quantum dynamical semigroup. Since relaxation rates are measurable quantities proposed constraints provides necessary physical condition for the Markovian generator to be physically legitimate. It is shown that the conjecture is supported by several well known examples of quantum semigroups including unital (doubly stochastic) evolution and semigroups derived in the weal coupling limit. It is strongly supported by numerical analysis (cf. Figure \ref{fig:ns1}). Interestingly, the conjecture has several important implications: it allows to provide universal constraints for spectra of quantum channels and provides necessary condition to decide whether a given channel $\Phi$ is consistent with Markovian evolution $\Phi = e^{\mathcal{L}}$. Note, that presented analysis may be immediately generalized for the time dependent case. Now, the evolution is generated by time-dependent generator $\mathcal{L}_t$. A question which attracted a lot attention recently -- is this evolution Markovian (cf. recent reviews \cite{NM1,NM2,NM3,NM4}). Now, having an access to local relaxation rates $\Gamma_\alpha(t)$ the constraint (\ref{CON}) provides necessary condition for Markovianity (defined via so called CP-divisibility \cite{RHP}). Hence, whenever local relaxation rates violate (\ref{CON}) the evolution is non-Markovian. In \cite{Sabrina} a hierarchy of $k$-divisibility ($k=1,\ldots,d$) was proposed --- it states that the propagator $V_{t,s}$ defined via $\Lambda_t = V_{t,s}\Lambda_s$ is $k$-positive.  Our original conjecture (\ref{CON}) strongly suggests that $k$-positivity is controlled by the following constraint $\Gamma \geq k \Gamma_\alpha$, which reproduces (\ref{CON}) if $k=d$. If this is true then purely mathematical property of the map ($k$-divisibility) can be decided in terms of purely physical quantities (local relaxation rates). This however needs further analysis.

\section*{Acknowledgements}

We would like to thank Y. Shikano and S. Ajisaka for their comments and discussion. We also thank K. \.Zyczkowski and S. Denisov for valuable discussions on random Lindblad generators. D. C. was supported by the Polish National Science Centre projects No. 2018/30/A/ST2/00837, respectively. {G. K. is supported in part by JSPS KAKENHI Grants No. 17K18107.}

\appendix

%\onecolumn

\section{No constraints for Classical Rates}\label{sec:NoBddCl}

Different from quantum cases, we observe no (non-trivial) constraints on the classical rates $\Gamma^{\rm cl}_k$.
In particular, any single relative rate $R^{\rm cl}_k := \Gamma^{\rm cl}_k/(\sum_l \Gamma^{\rm cl}_l)$ can be arbitrary close to `$1$'.
This can be shown by constructing a classical generator $K$ which possesses arbitrary positive classical rates $r_k \ge 0 \ (k=1,\ldots,d-1)$:
$$
K =
\left(
\begin{array}{cccccc}
-r_1 & r_2 & r_3 & \cdots & r_{d-1} & 0  \\
0 & - r_2 & 0 & \cdots & 0 & 0
\\
0 & 0 & -r_3 & \cdots& 0  & 0
\\
\vdots & \cdots & \vdots & \ddots & \vdots & \vdots \\
0 & 0 & 0 & \cdots& -r_{d-1}  & 0
\\
r_1 & 0 & \cdots & 0 & 0   & 0
\end{array}
\right).
$$
Clearly Kolmogorov conditions are satisfied and one can easily check that the eigenvalues of this matrix are $-r_j \ (j=1,\ldots,d-1)$ and $0$.

\section{Proof of Proposition \ref{PrCl}}\label{sec:PrPr1}

For the classical generator \eqref{L-class}, a straightforward calculation shows that $\LA^\ddag(E_{ii}) = \sum_j (t_{ij} - \delta_{ij} b_i )E_{jj} = \sum_j K_{ij} E_{jj}$, hence the diagonal elements satisfy $\frac{d}{dt} p_i = \tr E_{ii} \LA(\rho) = \tr \LA^\ddag (E_{ii}) \rho = \sum_j K_{ij} p_j$.
Similarly, one has $\LA(E_{ii}) = \sum_{j} K_{ji} E_{jj}$. Using this, one can check that the spectrum of $\LA$ consists of $d$ {\em classical} eigenvalues of the classical generator $K$:  $\lambda_0=0,\ell^{\rm cl}_1,\ldots,\ell^{\rm cl}_{d-1}$, and the remaining eigenvalues correspond to eigenvectors $E_{kl}$:
\begin{equation}\label{}
  \LA(E_{kl}) = - \frac 12(b_k + b_l) E_{kl} \ , \ \ \ (k \neq l) .
\end{equation}
Hence, one has {\em classical} rates $\Gamma^{\rm cl}_1,\ldots,\Gamma^{\rm cl}_{d-1}$,  and the remaining {\em quantum } rates
\begin{equation}\label{s:Gkl}
  \Gamma_{kl} = \frac 12 (b_k + b_l) , \ \ \ (k \neq l) .
\end{equation}
To prove Proposition \ref{PrCl}, we show
\begin{equation}\label{s:C1}
\Gamma^{\rm cl}_i \le \frac{1}{d}  \sum_{\beta=1}^{d^2-1} \Gamma_\beta  , \ \ \ i =1,\ldots,d-1 ,
\end{equation}
and
\begin{equation}\label{s:C2}
\Gamma_{kl} \le \frac{1}{d}  \sum_{\beta=1}^{d^2-1} \Gamma_\beta, \ \ \ k\neq l=1,\ldots,d.
\end{equation}
Reminding our convention that $t_{ii} = 0$,
\begin{equation*}\label{}
  \Gamma^{\rm cl} := \sum_{i=1}^{d-1} \Gamma^{\rm cl}_i  = - \tr K = - \sum_l (t_{ll} - \sum_{k=1}^d t_{kl}) = \sum_{k,l} t_{kl}.
\end{equation*}
Using (\ref{s:Gkl}), one has $\sum_{k \neq l} \Gamma_{kl} = (d-1) \sum_{k,l}t_{kl}$, and hence
\begin{equation*}\label{}
\Gamma:= \sum_{\beta=1}^{d^2-1} \Gamma_\beta = \Gamma^{\rm cl} + \sum_{k \neq l} \Gamma_{kl} = d \sum_{k,l} t_{kl} = d \Gamma^{\rm cl}.
\end{equation*}
Therefore, one finds
\begin{equation}\label{s:eq:ClGam}
 \frac{1}{d} \Gamma = \Gamma^{\rm cl} \geq  \Gamma^{\rm cl}_i , \ \ \ i =1,\ldots,d-1 ,
\end{equation}
which shows that  (\ref{s:C1})  is trivially satisfied.
Now, to prove (\ref{s:C2}) one needs to show
\begin{eqnarray*}\label{}
\frac{1}{d}  \Gamma  &=& \Gamma^{\rm cl} =  \sum_{i,j} t_{ij} \geq \frac 12 (b_k + b_l) \\
&=& \frac 1 2 \Big( \sum_{i} t_{ik} + \sum_{j} t_{jl} \Big)  , \ \ \ k\neq l=1,\ldots,d,
\end{eqnarray*}
which again is trivially satisfied by the positivity of $t_{ij}$.

\section{The bound is tight}\label{sec:TightBdd}

To see that the conjectured bound \eqref{CON} is tight, i.e., the constant $d$ is the best constant, one can simply construct a simple GKLS generator with which the equality in \eqref{CON} is attained.
The simplest one we have found is given by the generator \eqref{pure}.
Here, we show that it satisfies \eqref{CON} (the fact of which is covered by the general statement in Proposition \ref{prop:US}) and in particular that the equality is attained by an appropriate $\Sigma$.

By ordering the eigenvalues in ascending order: $s_1 \leq s_2 \leq \ldots \leq s_d$,  it is enough to prove
\begin{equation}\label{}
\sum_{k,l=1}^d \Gamma_{kl} \leq d \Gamma_{1d},
\end{equation}
that is,
\begin{equation}\label{1N}
  2 \sum_{k<l} (s_k - s_l)^2 \geq d (s_d - s_1)^2.
\end{equation}

\begin{Lemma}\label{s:lem:ElIneq} For any $x_1,x_2,\ldots,x_n \in [a,b]$ with real numbers $a < b$ and $n = 1,2,\ldots$, one has
$$
2 \sum_{i=1}^n ((x_i -a)^2 + (x_i -b)^2) + \sum_{i,j=1}^n (x_i - x_j)^2 \ge n (a-b)^2,
$$
and the equality holds iff $x_1 = \cdots = x_n = \frac{a+b}{2}$.
\end{Lemma}

[Proof of Lemma \ref{s:lem:ElIneq}] Define $f(x_1,\ldots,x_n) = 2 \sum_{i=1}^n ((x_i -a)^2 + (x_i -b)^2) + \sum_{i,j=1}^n (x_i - x_j)^2$. Since $f$ is continuous and differentiable on the compact region $[a,b]^n$, it has a minimum value and is attained by one of the extremum of $f$.
However, one has
\begin{eqnarray*}
\frac{\partial}{\partial x_i} f &=& 4(x_i-a) + 4(x_i-b) + 2 \sum_{k,l} (x_k-x_l)(\delta_{ik} - \delta_{il}) \\
&=& 4 ((2 + n) x_i - (a+b) - \sum_k x_k).
\end{eqnarray*}
Therefore, the only extremum is at $x_1 = x_2 = \cdots = x_n$, which turns out to be $\frac{a+b}{2}$ (from the condition $\frac{\partial}{\partial x_i} f = 0$).
Therefore, $f$ takes its minimum at this point and the substitution of $x_1 = x_2 = \cdots = x_n = \frac{a+b}{2}$ to $f$ gives $n (a-b)^2$.
\hfill $\square$

Applying this Lemma for $n= d-2$, $a = s_1,b = s_d$, and $x_1 = s_2, x_2 = s_3, \cdots, x_n = s_{d-1}$ implies (\ref{1N}).
Finally, notice that if we take $\Sigma$ with $s_2 = \ldots = s_{d-1} = \frac{s_1+s_d}{2} $ then $\sum_{i,j=1}^d \Gamma_{ij} = d \Gamma_{1d} $.
This shows that the equality in \eqref{CON} is attained.

\section{Derivation of Eq. \eqref{2}}\label{sec:Derivation(23)} % Check the number of equation in the main article which is labeled as \label{2}

Using the dual generator $\mathcal{L}^\ddag$ for GKLS generator \eqref{L}, which reads
\begin{equation}\label{s:DagL}
\LA^\ddag (X) = i [H, X] + \sum_k \gamma_k \left(L_k^\dagger X L_k -\frac 12\{L_k^\dagger L_k,X\}\right),
\end{equation}
one observes a useful identity \eqref{2}.
Indeed, applying $X^\dagger X$, $X^\dagger$ to $X$ in \eqref{s:DagL}, the left hand side of \eqref{2} is computed as
\begin{eqnarray*}
\sum_k \gamma_k (L_k^\dagger X^\dagger X L_k - L^\dagger_k X^\dagger L_k X - X^\dagger L^\dagger_k X L_k + X^\dagger L_k^\dagger L_k X),
\end{eqnarray*}
which coincides with the right hand side of \eqref{2}.

%&{} i [H, X^\dagger X] + \sum_k \gamma_k (L_k^\dagger X^\dagger X L_k -\frac 12\{L_k^\dagger L_k,X^\dagger X\}) \\
%&{}- \left(i [H, X^\dagger] + \sum_k \gamma_k (L_k^\dagger X^\dagger L_k -\frac 12\{L_k^\dagger L_k,X^\dagger\})\right)X \\
%&{} - X^\dagger \left(i [H, X] + \sum_k \gamma_k (L_k^\dagger X L_k -\frac 12\{L_k^\dagger L_k,X\})\right) \\

\section{Derivation of Eq. (31) }\label{sec:Derivation(31)} % Check the number of equation in the main article which is labeled as \label{eq:ID}

In this section, we derive Eq.~\eqref{eq:ID} by introducing the following general formula for the trace of GKLS generator \eqref{L}:
\begin{equation}\label{s:eq:trL}
\tr \LA = \sum_k \gamma_k (|\tr L_k|^2 - d \| L_k \|^2),
\end{equation}
where $\|A\|^2 = {\rm Tr}(A^\dagger A)$.
If one uses the normalized and traceless generator $L_k$ (indeed, without loss of generality, the trace part of $L_k$ can be renormalized to the Hamiltonian part), this can be simplified to
\begin{equation}\label{s:eq:trL2}
\tr \LA = -d \sum_k \gamma_k.
\end{equation}
This relation was previously shown by direct computations based on unitary operator basis in \cite{Wolf-Cirac} and matrix units in \cite{KAW}, respectively. Here, we give its simple derivation using the well-known super-operator representation of a $\mathcal{L}$ \cite{Watrous}

\begin{eqnarray*}
 \hat{\LA} &=& -i (H \otimes \oper - \oper \otimes H^T) + \sum_k \gamma_k L_k\otimes \overline{L_k}  \\ &-& \frac{1}{2} L^\dagger_k L_k\otimes \oper -  \frac{1}{2} \oper \otimes L^T_k \overline{L}_k .
\end{eqnarray*}
%\begin{equation}\label{s:eq:repTens}
%\Lambda  \to \hat{\Lambda} := (\tau_\Lambda)^\Gamma.
%\end{equation}
%Here, $\tau_\Lambda:= \sum_{i,j} \Lambda(\ketbra{i}{j}) \otimes \ketbra{i}{j}$ is the Choi-Jamiolkowski isomorphism \cite{ref:CJ} and $\Gamma$ operation is the involution defined by $\bracket{ij}{\tau^\Gamma kl}:= \bracket{ik}{\tau jl}$.
%Notic that \eqref{s:eq:repTens} is linear and trace preserving.
%Moreover, it is immediately seen that, for the left and right multiplication maps, $\LA_A(X):= AX$ and ${\cal R}_B(X) := X B$, one has
%\begin{eqnarray*}
%\widehat{\LA_A} = A \otimes \I, \ \widehat{{\cal R}_B} = \I \otimes B^T,
%\end{eqnarray*}
where $T$ is the transposition operation. Now the relation \eqref{s:eq:trL} is easily obtained by taking the trace operation and using the cyclic property of the trace. Finally, the facts that $\LA$ has $\ell_0 = 0$ eigenvalues and the complex eigenvalues of $\LA$ always appear as conjugate pairs shows
$$
\tr \LA = \sum_{\beta=0}^{d^2-1} \ell_\beta = - \sum_{\beta =1}^{d^2-1} \Gamma_\beta.
$$
Comparing this and \eqref{s:eq:trL2}, one gets Eq.~(31).

\section{General form of generator \eqref{ULU}}\label{sec:CovG-1}

In this section, we show that the covariant generators satisfying \eqref{ULU} has the form \eqref{COV} with \eqref{012}.
\begin{Lemma}
A linear map $\Phi$ satisfies
\begin{equation}\label{s:eq:CU}
 U_\mathbf{x} \Phi(X) U^\dagger_\mathbf{x} = \Phi(U_\mathbf{x} X U^\dagger_\mathbf{x}),
\end{equation}
for any $\mathbf{x} = (x_1,\ldots,x_d) \in \mathbb{R}^d$ if and only if it has a form
\begin{equation}\label{s:eq:repCU}
\Phi(X) = \sum_{i,j=1}^d a_{ij} E_{ij} X E_{ij}^\dagger + \sum_{i,j=1}^d b_{ij} E_{ii} X E_{jj}.
\end{equation}
\end{Lemma}
[Proof] If $\Phi$ has the form \eqref{s:eq:repCU}, then one simply verifies \eqref{s:eq:CU}. Assume that \eqref{s:eq:CU} is satisfied for any $\mathbf{x}$.
Then, one has
\begin{equation}\label{s:eq:idUxUd}
U_\mathbf{x} \Phi(E_{ij}) U_\mathbf{x}^\dagger = \Phi( U_\mathbf{x}  E_{ij} U_\mathbf{x}^\dagger ) = e^{i(x_i - x_j)} \Phi(E_{ij}) \ (i,j=1,\ldots,d).
\end{equation}
Letting $\Phi(E_{ij}) = \sum_{k,l=1}^d a_{ij;kl} E_{kl}$, this implies
$$
a_{ij;kl} = e^{-i(x_i-x_k)} e^{i(x_j - x_l)} a_{ij;kl}  \ (i,j,k,l = 1,\ldots,d).
$$
This can be true for any $\mathbf{x}$ if and only if $a_{ii;kl}$ has the form:
$$
a_{ii;kl} = c_{ik} \delta_{kl}
$$
and
$$
a_{ij;kl} = b_{ij} \delta_{ik}\delta_{jl} \ (i \neq j).
$$
In other words,
$$
\Phi(E_{ii}) = \sum_{k=1}^d c_{ik} E_{kk}
$$
and
$$
\Phi(E_{ij}) = b_{ij} E_{ij} \ (i \neq j).
$$
Therefore, the general form of $\Phi$ reads
\begin{eqnarray*}
&&\Phi (X) = \Phi(\sum_{i,j=1}^d \bracket{i}{X j} E_{ij}) \\
&=& \sum_{i,k=1}^d \bracket{i}{X i} c_{ik} E_{kk} +  \sum_{i \neq j=1}^d \bracket{i}{X j} b_{ij} E_{ij} \\
&=& \sum_{i,j=1}^d c_{ij}  E_{ji} X E_{ji}^\dagger + \sum_{i, j=1}^d b_{ij}  E_{ii} X E_{jj}.
\end{eqnarray*}
By adding an arbitrary $b_{ii}$ and letting $a_{ij} = c_{ij} - b_{ii} \delta_{ij}$, one gets the form \eqref{s:eq:repCU}. $\square$

Now, using the fact that any GKLS generator can be represented as

\begin{equation}\label{}
  \mathcal{L}(\rho) = -i[H,\rho] + \Phi(\rho) - \frac 12 \{\Phi^\ddag(\oper),\rho\} ,
\end{equation}
one finds that condition \eqref{ULU} implies that $H$ is diagonal $H=\sum_k h_k |k\>\<k|$ and hence one arrives at \eqref{COV}.

\section{Proof of Proposition \ref{prop:CG} }\label{sec:CovG-2} % In the main article, the proposition is labelled by {prop:CG}

Following the analysis of (\ref{L-class}) one easily finds the same set of {\rm classical} rates $\Gamma^{\rm cl}_\alpha$ and the rates $\Gamma_{kl}$ are simple modification of (\ref{s:Gkl}):
\begin{equation}\label{Gkl-new}
  \Gamma_{kl} = \frac 12 \Big( b_k + b_l +d_{kk} + d_{ll} - d_{kl} - d_{lk} \Big) , \ \ \ (k \neq l),
\end{equation}
and one has
\begin{equation}\label{}
\Gamma:= \sum_{\beta=1}^{d^2-1} \Gamma_\beta = d \Gamma^{\rm cl} + (d-1) \sum_i d_{ii}  -  \sum_{i \neq j} {\rm Re}\, d_{ij}.
\end{equation}
Since the positivity of the matrix $D = [d_{ij}]$ implies the positivity of every principal sub-matrix, ${\rm Re} \ d_{ij} \le |d_{ij}| \le \sqrt{d_{ii}d_{jj}} \le \frac{d_{ii}+d_{jj}}{2}$. Summing this over all $i \neq j$ gives
\begin{equation}\label{}
   (d-1) \sum_i d_{ii}  \geq   \sum_{i \neq j} {\rm Re}\, d_{ij} .
\end{equation}
It is therefore clear that
\begin{equation}\label{}
  \Gamma \geq d \Gamma^{\rm cl}_\alpha .
\end{equation}
Now, we prove  $\Gamma \geq d \Gamma_{kl}$. Without loosing generality we consider $k=1$ and $l=2$. Note that $\Gamma \geq d \Gamma_{12}$ is equivalent to the following inequality

\begin{equation}\label{xy}
  x + y \geq 0 ,
\end{equation}
where
\begin{equation}\label{x}
  x := d \sum_{i \neq j} t_{ij} - \frac d2 \left( \sum_{m \neq 1} t_{m1} + \sum_{n \neq 2} t_{n2} \right) ,
\end{equation}
and
\begin{equation}\label{y}
  y:= (d-1)  \sum_i d_{ii} -  \frac{1}{2} \sum_{i\neq j} (d_{ij} + d_{ji} ) - \frac d2 (d_{11} + d_{22}) + \frac d2 (d_{12}+d_{21}) .
\end{equation}
Now, since all $t_{ij} \geq 0$ one has $x\geq 0$ and hence to prove (\ref{xy}) it is enough to show that $y \geq 0$. Let us observe that
\begin{equation}\label{}
  y = \tr ( \mathbb{J} [A \circ D] ) ,
\end{equation}
where $A \circ D$ denotes Hadamard product, $\mathbb{J}_{ij}=1$, and
\begin{equation}\label{A}
  A = \left( \begin{array}{ccccc} \frac d2 -1  & \frac d2 -1 & -1 & \ldots & -1 \\  \frac d2 -1 & \frac d2 -1 &  -1 &\ldots  &-1 \\-1 & -1 & d-1 & \ldots  & -1 \\ \vdots & \vdots &\ddots &\vdots &\vdots  \\ -1 & -1 & \ldots & d-1 & -1 \\-1 & -1 &\ldots &  -1 & d-1  \end{array} \right) .
\end{equation}
The matrix $\mathbb{J}_{ij}$ is positive definite. One finds that the eignevalues of the matrix $A$ read: $\{0,0,d,\ldots,d\}$ which proves that $A$ is positive definite as well. Since $D \geq 0$ one has $A \circ D \geq 0$. Finally $y \geq 0$ since the trace of $\mathbb{J} [A \circ D]$ has to be positive. This completes the proof of Proposition \ref{prop:CG}.

\end{document}